\documentclass{osa-article}
\journal{osajournal}
\articletype{Research Article}
\usepackage{lineno}

\begin{document}
\title{Learned holographic light transport}
\author{Koray Kavakl\i,\authormark{1} Hakan Urey,\authormark{1} and Kaan Ak\c{s}it\authormark{2,*}}
\address{\authormark{1}Department of Electrical and Electronics Engineering, Ko\c{c} University, Istanbul, Turkey\\
\authormark{2}Department of Computer Science, University College London, London, UK}
\email{\authormark{*}k.aksit@ucl.ac.uk}
\homepage{https://kaanaksit.com}

\begin{abstract}
Computer-Generated Holography (CGH) algorithms often fall short in matching simulations with results from a physical holographic display.
Our work addresses this mismatch by learning the holographic light transport in holographic displays.
Using a camera and a holographic display, we capture the image reconstructions of optimized holograms that rely on ideal simulations to generate a dataset.
Inspired by the ideal simulations, we learn a complex-valued convolution kernel that can propagate given holograms to captured photographs in our dataset.
Our method can dramatically improve simulation accuracy and image quality in holographic displays while paving the way for physically informed learning approaches.
\end{abstract}

\section{Introduction}
The future of human-computer interactions~\cite{orlosky2021telelife} demands technologies that can display life-like three-dimensional visuals.
An emerging trend, Computer-Generated Holography (CGH), promises to deliver such realistic visuals in the next-generation displays~\cite{koulieris2019near}.
However, CGH algorithms often fall short of achieving high image quality in real life.

The traditional CGH algorithms such as Gerchberg-Saxton method~\cite{yang1994gerchberg} or recently trending approaches such as Stochastic Gradient (SGD) based differentiable methods~\cite{chen2021multi,peng2020neural,chakravarthula2020learned} can deliver an outstanding image quality in the simulation environments.
However, in an actual holographic display with phase-only modulation, holograms optimized or learned using these ideal holographic light transport models often fail to deliver the same image quality.
Identifying causes of mismatch and bridging the gap between the image qualities of simulations and actual experiments are growing scientific research trends in the holography community.

The traditional solutions~\cite{li2019progress} to address the mismatch aims to find complex residual values that can be added as a regularization term to the ideal holographic light transport~\cite{chakravarthula2020learned} or the complex hologram~\cite{peng2020neural}.
These techniques to regularize holographic image reconstruction models~\cite{peng2020neural,chakravarthula2020learned} are powerful and effective in practice.
In the meantime, researchers have also garnered interest to learn the hologram generation process using deep learning~\cite{peng2020neural,wu2021high}.
However, their proposed solutions often yield highly complex algorithmic structures and sometimes require a physically demanding experimentation routine.
These complex algorithmic structures involve learning components such as Generative Adversarial Networks (GANs) that are not straightforward in tuning and training~\cite{chakravarthula2020learned}, multi-layer perceptrons that model the nonlinear response of an SLM, which may carry lesser semantic meaning for an optical scientist~\cite{peng2020neural} or characterizing aberrations with Zernike polynomials that requires careful experimentation~\cite{peng2020neural,zhao2017multi,zhang2017calibration,krasin2021holographic,xun2004phase}.
We ask ourselves if the demand in experimentation load and complex nature of algorithms can be avoided while optical scientists can get more hints towards understanding imperfections in actual holographic displays.
With that question in mind, we aim for deriving a new and refined CGH algorithm to improve image quality in actual holographic displays.

This work argues that a tailored holographic light transport model for a target holographic display can account for optical aberrations and bridge the gap between simulations and actual holographic displays.
We also argue that such a model can avoid intensive experimentation requirements in display calibration.
For this purpose, we propose to learn a single complex-valued point spread function that helps us to propagate input phase-only holograms to the target image plane.
Thus, our holographic light transport model convolves an input phase-only hologram with a learned complex-valued point spread function to get to the physically accurate image reconstructions in simulations for a target holographic display.
The learning process involves comparing image reconstruction in simulations against experiments using a camera with an actual holographic display.
Like any other learning process, we must have a set of data composed of input phase-only holograms and their corresponding image reconstructions in an actual holographic display.
We collect such a dataset from our proof-of-concept display prototype using a camera and an ideal holographic light transport based hologram optimization method that is fully differentiable.
We show that our learned holographic light transport can dramatically improve simulation accuracy and final image quality in our holographic display.
Our key contributions are summarized as follows:

\begin{itemize}
    \item Learned holographic light transport. We propose a learned approach for \textit{holographic light transport} to bridge the gap between simulations and experimentation. Our method learns a single complex convolutional kernel to reconstruct images in simulation similar to the real experiments. Our implementation is fully differentiable.  We show that image quality results from an actual holographic display can be enhanced with our method while the simulations become highly accurate.
    \item Holographic dataset from a proof-of-concept holographic display. In order to be able to train and derive a single complex convolutional kernel, we build a phase-only holographic display. Then, we capture a series of photographs of holographic image reconstructions resulting from holograms optimized using the ideal holographic light transport.
\end{itemize}

In the following sections, we will first introduce a standard ideal holographic light transport model. Then, we will provide the details of our experimental setup. Finally, we will introduce our technique in learning and provide quantitative results of our method while comparing it against the ideal case.

\section{Optimizing holograms with ideal holographic light transport}
\label{sec:ideal_light}
The topic of light transport plays a crucial role in formulating the basis of various domains including traditional computer graphics~\cite{vorba2014line}, architecture~\cite{ayoub2020review}, biomedical imaging~\cite{jonsson2020multi}, non-line-of-sight imaging~\cite{reza2019phasor}, three-dimensional printing~\cite{rittig2021neural}, visible light communications~\cite{corbellini2014connecting}, holographic recording~\cite{jang2020design}, computational displays~\cite{akcsit2020patch}, eye prescription correction~\cite{chakravarthula2018focusar}, eye-gaze tracking~\cite{li2020optical}, ophthalmology~\cite{aydindougan2021applications} and many more.
Although we cover only display technologies in this work, an accurate representation method of light transport can potentially pave the way towards enhancements in many other highlighted applications.

Light transport models used in CGH bases on Rayleigh-Sommerfeld diffraction integrals~\cite{heurtley1973scalar}. 
This diffraction integral's first solution, the Huygens-Fresnel principle, is expressed as follows:
\begin{equation}
u(x,y)=\frac{1}{j\lambda} \int\int u_0(x,y)\frac{e^{jkr}}{r}cos(\theta)dxdy,
\end{equation}
where resultant field, $u(x,y)$, is calculated by integrating over every point across hologram plane in XY axes, $u_0(x,y)$ represents the optical field in the hologram plane for every point across XY axes, $r$ represents the optical path between a selected point in hologram plane and a selected point in target plane, $\theta$ represents the angle between these points, $k$ represents the wavenumber ($\frac{2\pi}{\lambda}$) and $\lambda$ represents the wavelength of light.
In this model, optical fields, $u_0(x,y)$ and $u(x,y)$, are represented with a complex value,
\begin{equation}
u_0(x,y)=A(x,y)e^{j\phi(x,y)},
\end{equation}
where A represents the spatial distribution of amplitude and $\phi$ represents the spatial distribution of phase across a hologram plane.
To simplify our description, we can express the Huygens-Fresnel principle as a superposition of diverging spherical waves originating from a hologram~\cite{goodman2005introduction}. 
Perhaps one can also think of the Huygens-Fresnel principle as stamping a complex point-spread function on a target image plane for each point of a hologram while weighting each stamp with its amplitude and phase from its origin.

Calculating Huygens-Fresnel approximation by visiting each point on a hologram one by one would consume a large computation and power budget while being slow in processing.
Common approaches in the literature~\cite{matsushima2009band,zhang2020band,zhang2020adaptive} dedicated to near fields (e.g., short distances like 10 cm or half a meter) formulates this integral as a convolution operation with a single complex kernel.
Hence, the common approaches~\cite{sypek1995light} can be expressed as 
\begin{equation}
\begin{split}
u(x,y)=u_0(x,y) * h(x,y) \\ =\mathcal{F}^{-1}(\mathcal{F}(u_0(x,y)) \mathcal{F}(h(x,y))) \\ =\mathcal{F}^{-1}(U_0(f_x,f_y) H(f_x,f_y)),
\end{split}
\label{equ:holographic_light_transport}
\end{equation}
where h represents a spatially varying complex convolution kernel.
The value of the complex kernel, h, is typically expressed as
\begin{equation}
h(x,y)=\frac{e^{jkz}}{j\lambda z} e^{\frac{jk}{2z} (x^2+y^2)},
\end{equation}
where z represents the distance between a hologram plane and a target image plane.
This ideal model is implemented in a differentiable fashion (refer to \href{https://github.com/kunguz/odak/blob/12a3cc90fca58c2e2aeb86e18382e18423362bdb/odak/learn/wave/classical.py#L81-L114}{odak.learn.wave.classical L81-114}) in our fundamental library for optical sciences~\cite{kaan_aksit_2021_5136406}.
The same library hosts differentiable models of various light transport approximations (refer to \href{https://github.com/kunguz/odak/blob/12a3cc90fca58c2e2aeb86e18382e18423362bdb/odak/learn/wave/classical.py#L8-L53}{odak.learn.wave.classical L8-53}).

Now that we have established an ideal holographic light transport model as in Equation~\ref{equ:holographic_light_transport}.
We can use this holographic light transport model as a forward model that propagates light from a hologram to a target plane.
As mentioned earlier, since this model is implemented in code using a modern machine learning library, PyTorch~\cite{paszke2017automatic}, we take advantage of the fact that modern machine learning libraries are capable of automatically differentiating provided functions.
Differentiation helps to calculate the complex gradient of our forward model's error.
In simple terms, for each input phase-only hologram, the resulting image reconstruction can be calculated, and the impact of changing phase values on image reconstruction can be precisely estimated using gradients.
This fact helps an optimizer to have meaningful modifications on phase values of a phase-only hologram at the each optimization step.

In order to fully realize the described optimization, a loss function is required.
For this purpose, we define a loss function, $L$, using least squared error between a reconstructed image at a target plane, u(x,y), and a target image, t(x,y),
\begin{equation}
L=(u(x,y)-t(x,y))^2.
\label{equ:loss}
\end{equation}
Note that the loss function described here is the simplest case, and we leave customization of this loss function to meet the application's demands as a future discussion.
In addition to a loss function, we would require an optimizer to optimize our phase-only holograms for various targets.
We choose to use a Stochastic Gradient Descent based optimization method~\cite{kingma2014adam,loshchilov2017decoupled} with a learning rate of $0.1$.
We ran our optimizer using our ideal forward model for  200 iterations at each hologram calculation.
Our hologram optimization method (refer to \href{https://github.com/kunguz/odak/blob/12a3cc90fca58c2e2aeb86e18382e18423362bdb/odak/learn/wave/classical.py#L232-L298 L232-298}) is distributed as a part of our fundamental library for optical sciences~\cite{kaan_aksit_2021_5136406}.
We also provide examples at \href{https://github.com/kunguz/odak/blob/master/test/test_learn_sgd.py}{odak.test.learn\_sgd} for using the optimization method within our library.

Using the described ideal holographic light transport and hologram optimization methodology, we calculate phase-only holograms of target images from DIV2K dataset~\cite{Ignatov_2018_ECCV_Workshops}.
We resize images in DIV2K dataset to $1920x1080$ to match the size of our SLM.
We also convert those images to monochrome by taking an average across three color channels.
Note that all these images at DIV2K dataset are used only in training (finding the holographic light transport kernel, not in test cases).
The calculated image reconstructions perfectly matching the target images in simulations, however they have to be tested against photographs captured from an actual holographic display. 
Thus, we will explain how we build a proof-of-concept holographic display in the next section before explaining our final methodology to improve visual quality and realism.

\section{Proof-of-concept holographic display}
We build a proof-of-concept holographic display to assess image quality of our hologram optimization methods that uses ideal holographic light transport.
We will introduce our learned holographic light transport in the next section, we will also use the same proof-of-concept holographic display to assess image quality of our methodology.

\begin{figure}[h!]
\centering
\includegraphics[width=\textwidth]{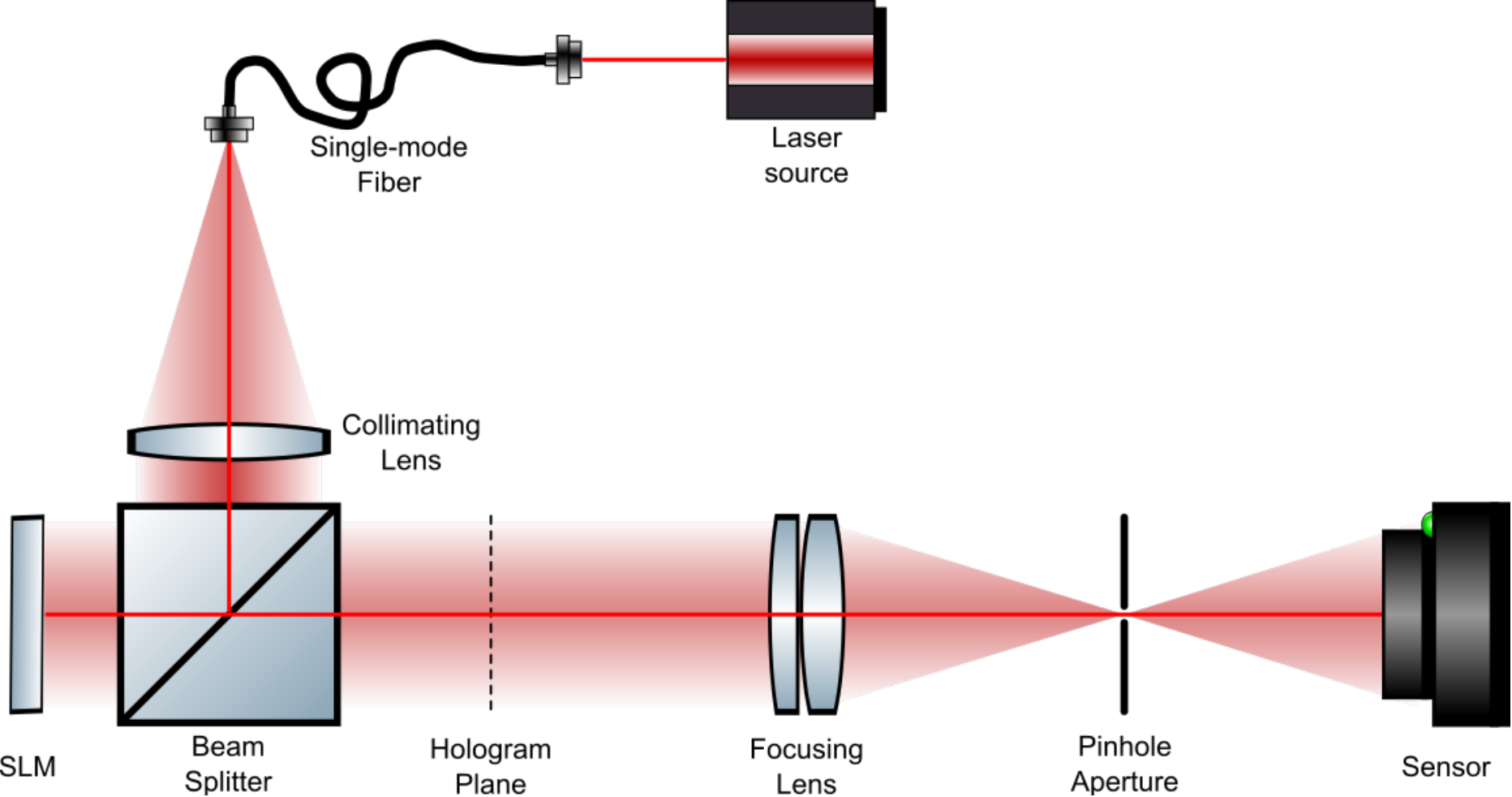}
\caption{Schematic diagram of our proof-of-concept holographic display prototype used in our experimental setup.}
\label{fig:display_prototype}
\end{figure}

The optical layout of our proof-of-concept holographic display is represented in Figure~\ref{fig:display_prototype}.
Following the light from its source, the optical assembly of our proof-of-concept holographic display uses a multi-wavelength laser light source, LASOS MCS4.
However, for our experimentation, we only rely on the working wavelength of 515~nm.
A Thorlabs LB1945-A bi-convex lens with a 200 mm focal length lens collimates the output beam of our laser light source.
The collimated beam goes through a wire grid linear polarizer, Thorlabs LPVISE100-A, to maintain a polarization aligned with our phase-only Spatial Light Modulator's fast axis (SLM).
The linearly polarized collimated beam bounces off an anti-reflection coated Pellicle beamsplitter, Thorlabs BP245B1, towards our $0.90$ degrees tilted phase-only SLM, Holoeye Pluto 2.0 (tilted half order). 
To avoid undiffracted light, we add a horizontal grating to the displayed holograms on our SLM. 
The horizontally grated hologram, $u_0'$ can be calculated as 
\begin{equation}
  u_0'(x,y) =
  \begin{cases}
            e^{-j(\phi(x,y) + \pi)} & \text{for $x=$ odd} \\
            e^{-j\phi(x,y)}         & \text{for $x=$ even} \\
  \end{cases}
  \label{equ:horizontal_grating}
\end{equation}
where $\phi$, the original phase of $u_0$, is modified. 
This way, we steer the location of the reconstructed image in space away from undiffracted light. 
The tilt angle of our SLM calculated using the diffraction equation formulated as
\begin{equation}
    m\lambda = {\Delta}a sin(\theta),
\end{equation}
where m is the half-order (0.5), ${\Delta}a$ is pixel pitch of a SLM and the $\theta$ is the angular location of the grated hologram plane.
For our system, $\theta$ is calculated as $1.80^{\circ}$. 
So the required tilt angle for the SLM is $\frac{\theta}{2} \approx 0.90^{\circ}$. 
In the rest of the setup, the phase-modulated beam goes through the Pellicle beamsplitter.
In the next stage, the beam passes focusing lenses, a combination of Thorlabs LA1908-A and LB1056-A.
A pinhole aperture, Thorlabs SM1D12, follows the lenses at the focal distance of the focusing lenses to avoid undiffracted light.
We capture the image reconstructions of our hologram dataset optimized using ideal holographic light transport from our setup with a lensless image sensor, Point Grey GS3-U3-23S6M-C USB 3.0.
For each captured image reconstruction, we applied homography correction for the captures, so that we can compare it against a ground-truth image or a simulated reconstruction.
The holograms in our work are always reconstructed for a target image plane at 7~cm away from our proof-of-concept holographic display.

\section{Learned holographic light transport}
\label{sec:learned_light}
\begin{figure}[ht!]
\centering
\includegraphics[width=\textwidth]{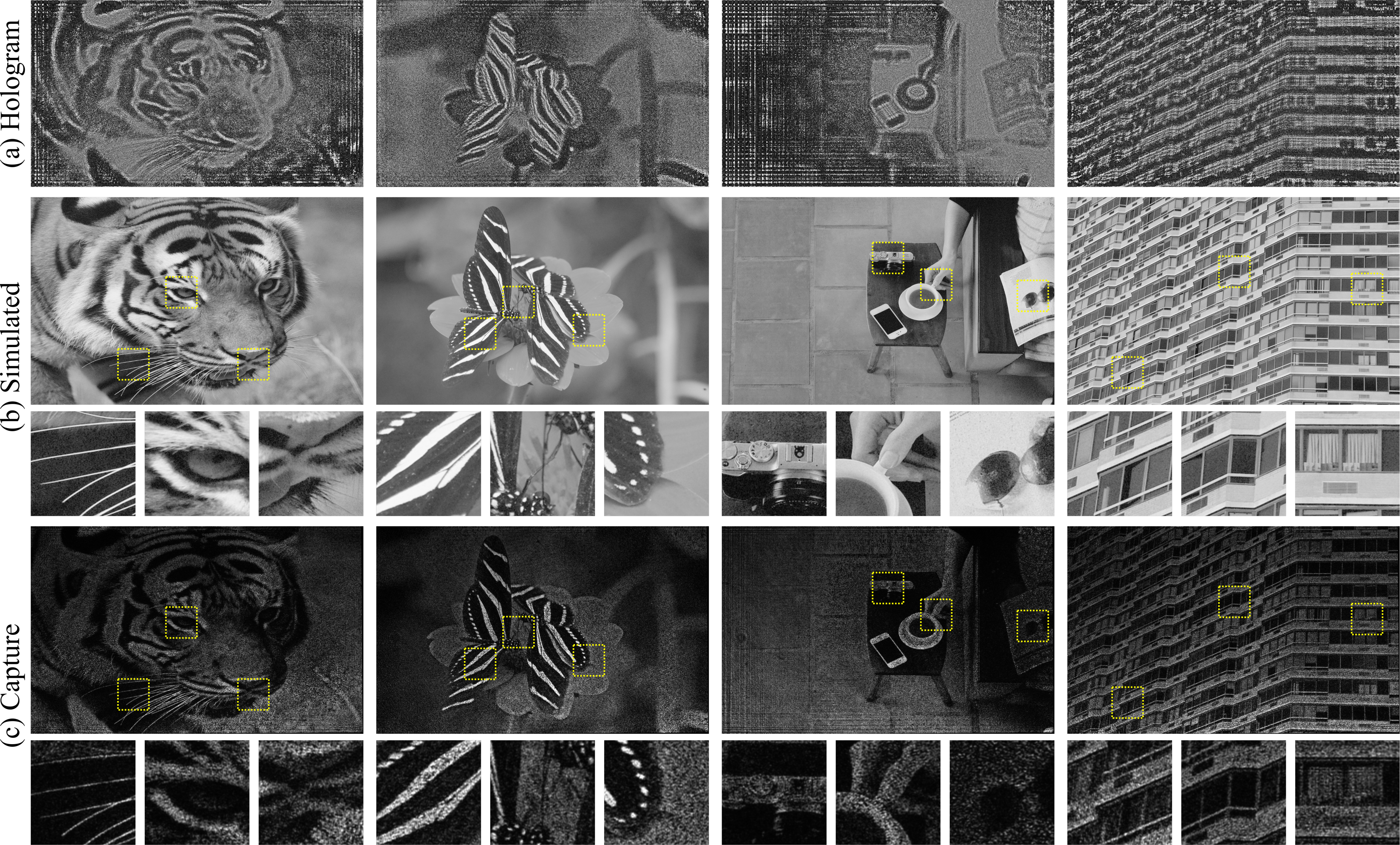}
\caption{Mismatch between simulated and experimental results when using ideal holographic light transport. 
For a given (a) phase-only hologram, A simulated result can provide (b) a perfect image reconstruction, while the same hologram in (c) a real holographic display fail in achieving such image reconstructions as we show in Dataset 1 (Ref. \cite{kavakli_2021}).}
\label{fig:sample_ideal_results}
\end{figure}

We provide sample photographs showing image reconstructions captured from our proof-of-concept holographic display in Figure~\ref{fig:sample_ideal_results}.
These photographs are a result of holograms optimized using the ideal holographic light transport model.
We also provide input holograms and their simulated results for comparison.
The visual mismatch between photographs and simulated results provides a good understanding of the image quality issues discussed earlier.

\begin{figure}[ht!]
\centering
\includegraphics[width=12cm]{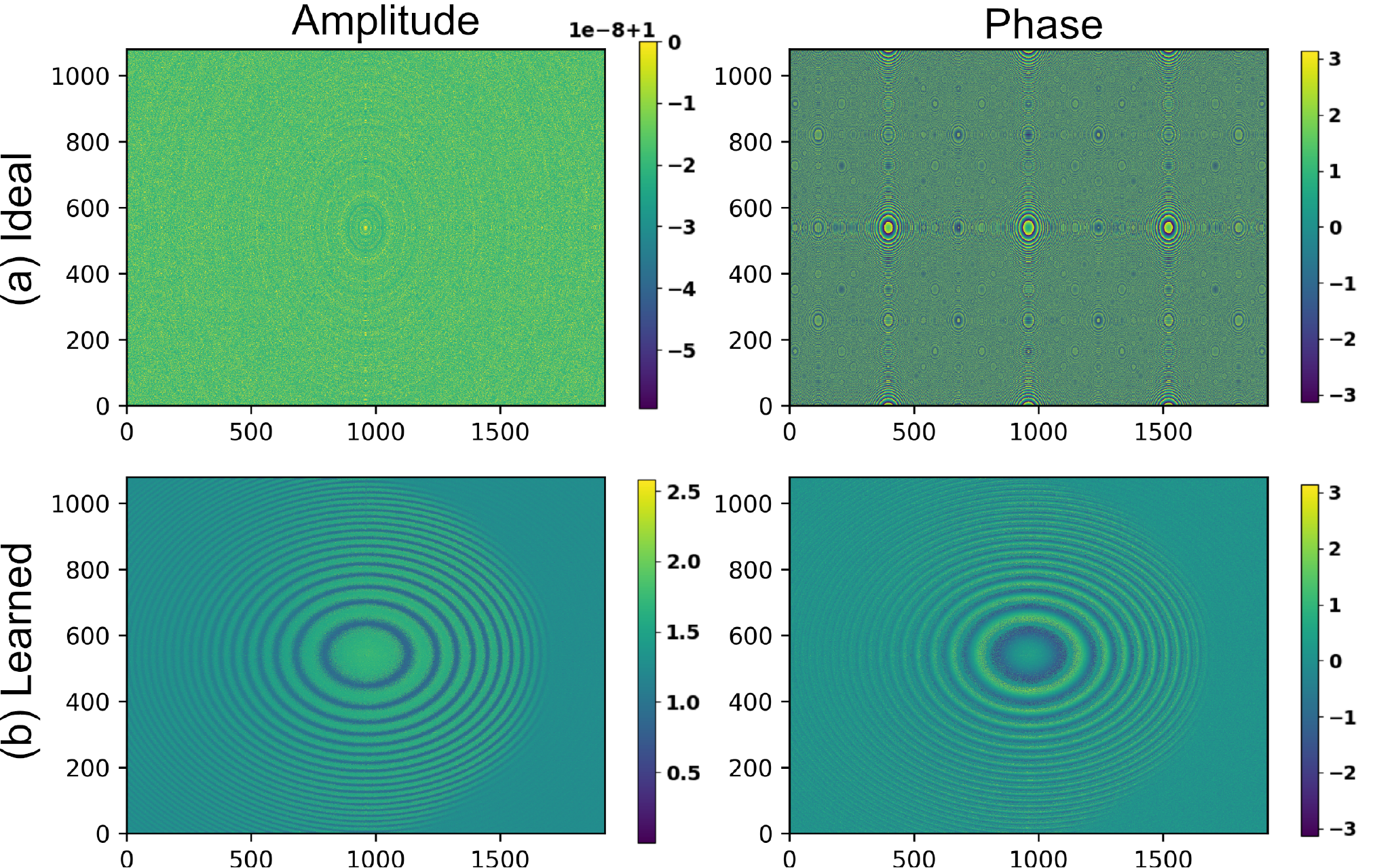}
\caption{A phase and amplitude comparison between complex kernels used in (a) ideal holographic light transport and (b) learned holographic light transport.}
\label{fig:ideal_vs_learned_kernels}
\end{figure}

To combat this mismatch illustrated in Figure~\ref{fig:sample_ideal_results}, we take advantage of our dataset of photographs from the proof-of-concept prototype and their corresponding optimized holograms that used the ideal holographic light transport model (Dataset 1 \cite{kavakli_2021}).
With a Stochastic Gradient Descent based optimization method~\cite{kingma2014adam,loshchilov2017decoupled} and a learning rate of 0.002, we set to learn a complex kernel, $h_l(x,y)$ using the loss function at Equation~\ref{equ:loss} that will replace the original $h(x,y)$ from the ideal case.
This newly optimized $h_l(x,y)$ can be best described as a transfer function that takes an ideal input hologram and provides an image reconstruction similar to the captured photographs in our dataset.
The code base of our learning process follows the same optimization described in Section~\ref{sec:ideal_light} (refer to \href{https://github.com/complight/realistic_holography/blob/3815a1d349adf6e0059eb6dba1b770e104ed8637/optics.py#L87-L137}{realistic$\_$holography:optics L87-L137}).
The phase and amplitude of the learned complex kernel, $h_l(x,y)$, and the ideal complex kernel, $h(x,y)$, are provided in Figure~\ref{fig:ideal_vs_learned_kernels} for comparison.

\section{Evaluation}
Now that we have a learned transfer function, $h_l(x,y)$, shown in Figure~\ref{fig:ideal_vs_learned_kernels}, we look into how this kernel representing the learned holographic light transport can help us to optimize new holograms.
Assume that the learned kernel is more realistic than the original ideal kernel. 
In that hypothesis, the optimized holograms should lead to image reconstruction results better in terms of image quality in the experimental case.
Meantime, we should also expect that the mismatch between simulations and experiment cases to be mitigated.
We challenge these assumptions by optimizing holograms using $h_l(x,y)$ instead of $h(x,y)$. 
In our exploration for optimizing holograms using the learned holographic light transport, we rely on the same process described in Section~\ref{sec:ideal_light} (refer to \href{https://github.com/complight/realistic_holography/blob/3815a1d349adf6e0059eb6dba1b770e104ed8637/optics.py#L45-L85}{realistic\_holography:optics L45-85}).

\begin{figure}[ht!]
\centering
\includegraphics[width=\textwidth]{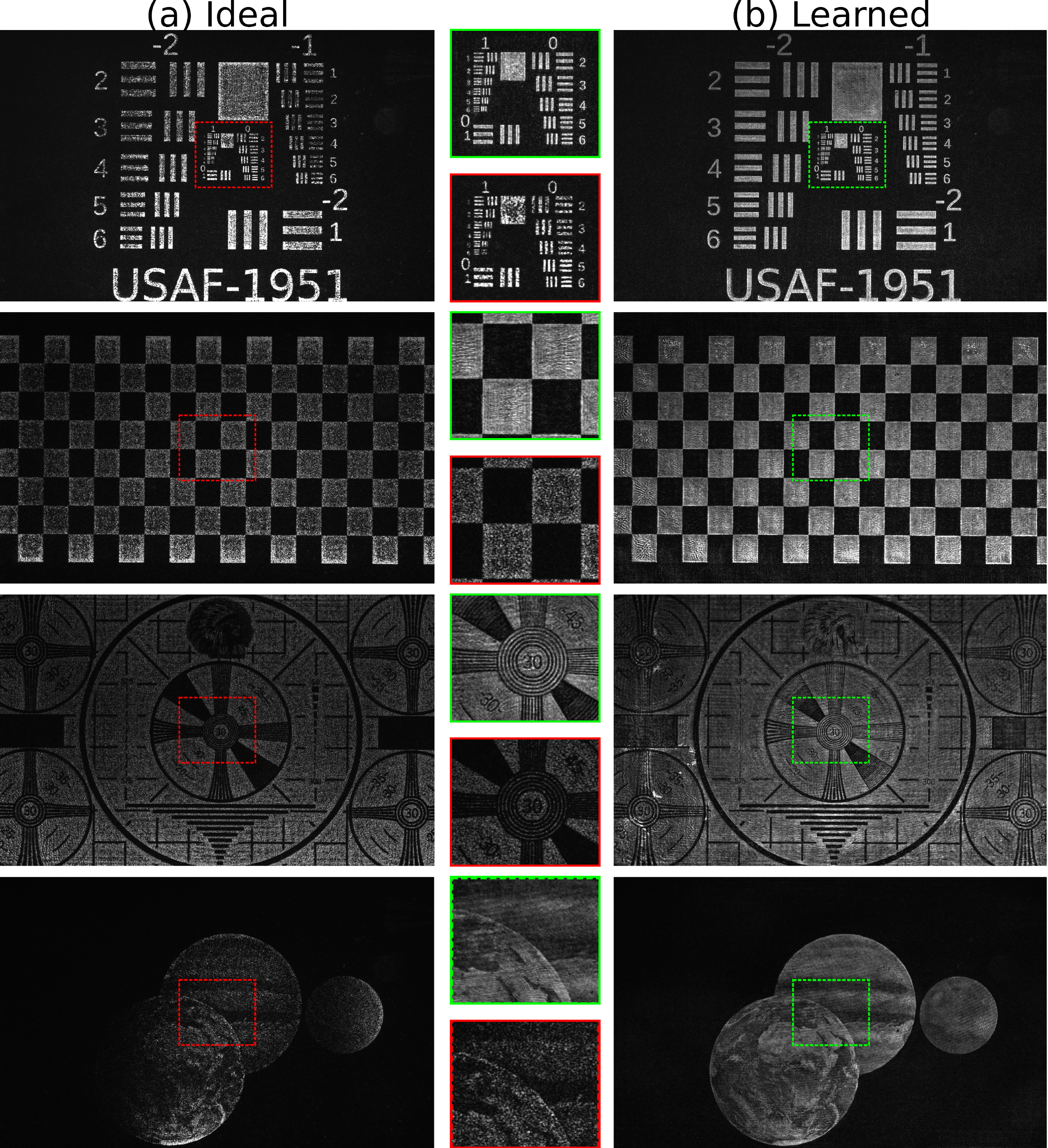}
\caption{A visual comparison between (a) ideal holographic light transport and (b) learned holographic light transport in reconstructing images. Both of the photographs are captured with optimized holograms using corresponding holographic light transport models and our proof-of-concept prototype. Note that target image at both cases are not used in our training set (DIV2K~\cite{Ignatov_2018_ECCV_Workshops}).}
\label{fig:ideal_vs_learned}
\end{figure}

\paragraph{Image quality.} We provided a visual comparison between holograms generated using the ideal holographic light transport and learned holographic light transport in Figure~\ref{fig:ideal_vs_learned}.
The visual quality of the reconstructed images in our proof-of-concept holographic display using the learned holographic light transport shows a significant improvement over the ideal case.
We believe this is because imperfections in our proof-of-concept holographic display are accounted for in our transfer function.
We kindly invite the readers to observe the visual difference between the ideal transfer function and the learned transfer function provided in Figure~\ref{fig:ideal_vs_learned_kernels}.
Please note the asymmetry in the learned kernel, which does not exist in the case of an ideal kernel.

\begin{figure}[ht!]
\centering
\includegraphics[width=\textwidth]{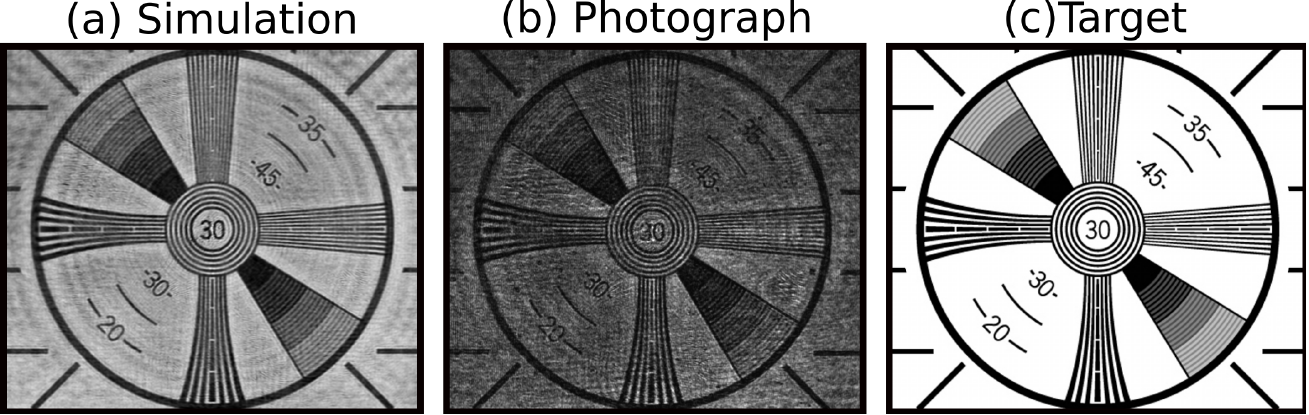}
\caption{Learned simulation (a) versus real photograph (b). Ideal light transport based hologram optimization estimates unrealistic results in simulation. On the other hand, for a given target image (c), simulations based on learned holographic light transport closely resembles the experimental results.}
\label{fig:learned_simulation_vs_real}
\end{figure}

\paragraph{The mismatch between simulations and experiments.} Our learned holographic light transport can approximate a transfer function of our proof-of-concept display accurately (Training L2 loss: 0.0028 and test loss: 0.0034 -- learned reconstruction versus captured ground truth images -- note that images are normalized between zero and one).
We compare image reconstructions from our simulations with our experimental results from our proof-of-concept holographic display to provide evidence that this is the case.
This comparison is sampled in Figure~\ref{fig:learned_simulation_vs_real}.
Our simulations' brightness and contrast levels with learned holographic light transport do not truly match our photographs from our experimentation.
However, the spatial content in experimental cases resembles the simulated reconstructions closely and even giving us an excellent hint about what to expect in terms of visual quality from a given holographic display.
If further tweaking is needed, the brightness mismatch in the ideal and learned cases can be improved by following a manual calibration routine.
In the supplementary documentation of work by Choi et al.~\cite{choi2021optimizing}, curious readers can find highly detailed documentation on minimizing the average difference between simulation and a physical prototype by adjusting laser power and exposure time.
We have not conducted such a calibration for this work, as we wanted to show the improvement over an uncalibrated system.

\paragraph{What did we learn from the learned kernel?}
The holographic light transport kernel learned within this work (see Figure~\ref{fig:ideal_vs_learned_kernels}) indicates that the phase and amplitude behaviour of our physical light source is not homogenous in terms of angular emission.
The readers may observe this fact by carefully checking the asymmetry of the kernel in Figure~\ref{fig:ideal_vs_learned_kernels}.
The amplitude values are far greater than the ideal kernel, thus suggesting that to get to brighter images, the hologram optimization has to consider this correlation.
This fact can be observed in Figure~\ref{fig:ideal_vs_learned} as the dynamic range, and brightness levels are better preserved in the learned method.
Note that the learned kernel is the point-spread function of the given holographic display.
Thus, resolution characteristics of the holographic display can also be analyzed in the future by studying the limits of a learned point-spread function.
Finally, note that a single kernel can only capture a global mean of a general trend in a holographic display. 
We will discuss how to improve our learned method in the future in the final paragraph of this section.

\paragraph{Comparison with the state of the art.}
The leading state-of-the-art methods~\cite{peng2020neural,chakravarthula2020learned} that bridge the gap between simulations and physical holographic displays consists of convolutional neural networks.
Specifically, the work by Peng et al.~\cite{peng2020neural} relies on more than eight million parameters to tune in a training process of neural networks.
Many parameters and layers are needed to efficiently realize the correlation between a hologram and a final reconstructed image.
Otherwise, the connections between the pixels of a hologram and a final reconstructed image may not be fully identified (locality issue).
This locality issue arises from the fact that such models use small kernel sizes.
In contrast, our work decreases this number of tunable parameters to half, four million parameters (2x1080x1920 -- amplitude and phase), while relying on kernel sizes that is the same as an input image which avoids locality issue.
The work by Maimone et al.~\cite{maimone2017holographic} uses one-dimensional separable functions for reducing the memory footprint in classical CGH~\cite{maimone2017holographic} for ideal complex forms such as quadratic phase functions.
Drawing inspiration from that work, we speculate that further reduction may be possible by storing a parametric form of a learned complex kernel.

The readers of our work may ask if our approach is the solution that could provide the most remarkable accuracy in bridging the gap between simulations and experiments in holography.
We followed a similar approach to the classical model, where a single convolutional kernel formulates the transfer function of light transport.
Hence, our approach is accurate as long as a single kernel is reliable to describe the light transport. 
To improve accuracy further and to have one-to-one matching simulations in the future, we speculate that approaches with spatially varying convolutional kernels can provide more capacity to accommodate for a genuinely realistic simulation.
On the other hand, we learn the light transport between a hologram and an image plane. 
Approaches that provide three-dimensional image reconstructions in CGH require a transfer function representing the relationship between a single hologram and multiple image planes.
In our approach, we have to learn the kernel for each plane using a set of images. 
In the future, a complete form of our approach can potentially be derived where a data set with diverse image reconstruction distances helping to learn a parametric light transport rather than per plane learning.
Our work does an excellent job in capturing optical aberrations and imperfections of a holographic display.
Our work can be best described as the simplest form of improving realism in CGH algorithms without dealing with complex experimentation or complex algorithmic approaches.

\section{Conclusion}
Holographic displays often require tedious effort to optimize holograms for the best possible image quality.
We propose a new learned method to address this issue with holographic displays in a simple way.
The core of our approach is in a learning procedure that allows one to approximate an accurate holographic light propagation model for a given actual holographic display.
With this approach, we can optimize holograms that can dramatically improve image quality concerning a typical ideal holographic light transport model.
Our method, in turn, enables a simple yet effective method that does not suffer from the overhead of deriving complex algorithmic approaches while paving the way towards physically informed learning approaches in the holography domain.

\begin{backmatter}
\bmsection{Acknowledgments,}
The authors thank the anonymous reviewers for their useful feedback. The authors also thank Oliver Kingshott and Duygu Ceylan for the fruitful and inspiring discussions improving the outcome of this research, and Selim Ölçer for helping with the fiber alignment of laser light source in the proof-of-concept display prototype.

\bmsection{Disclosures,}
The authors declare no conflicts of interest.

\bmsection{Data Availability Statement} The code base discussed in Section~\ref{sec:ideal_light} and~\ref{sec:learned_light} is readily available in the \href{https://github.com/complight/realistic_holography}{ Github:complight/realistic$\_$holography}. The generated dataset of this work is also available in the \href{https://figshare.com/s/df4c8ff98fb271a3eba7}{Dataset: Phase-only holograms and photographs}.
\end{backmatter}

\bibliography{sample}

\begin{thebibliography}{10}
\newcommand{\enquote}[1]{``#1''}

\bibitem{orlosky2021telelife}
J.~Orlosky, M.~Sra, K.~Bekta{\c{s}}, H.~Peng, J.~Kim, N.~Kos'~myna,
  T.~Hollerer, A.~Steed, K.~Kiyokawa, and K.~Ak{\c{s}}it, \enquote{Telelife:
  The future of remote living,} {\protect\JournalTitle{arXiv preprint
  arXiv:2107.02965}}  (2021).

\bibitem{koulieris2019near}
G.~A. Koulieris, K.~Ak{\c{s}}it, M.~Stengel, R.~K. Mantiuk, K.~Mania, and
  C.~Richardt, \enquote{Near-eye display and tracking technologies for virtual
  and augmented reality,} in \emph{Computer Graphics Forum,}  vol.~38 (Wiley
  Online Library, 2019), pp. 493--519.

\bibitem{yang1994gerchberg}
G.-z. Yang, B.-z. Dong, B.-y. Gu, J.-y. Zhuang, and O.~K. Ersoy,
  \enquote{Gerchberg--saxton and yang--gu algorithms for phase retrieval in a
  nonunitary transform system: a comparison,} {\protect\JournalTitle{Applied
  optics}} \textbf{33}, 209--218 (1994).

\bibitem{chen2021multi}
C.~Chen, B.~Lee, N.-N. Li, M.~Chae, D.~Wang, Q.-H. Wang, and B.~Lee,
  \enquote{Multi-depth hologram generation using stochastic gradient descent
  algorithm with complex loss function,} {\protect\JournalTitle{Optics
  Express}} \textbf{29}, 15089--15103 (2021).

\bibitem{peng2020neural}
Y.~Peng, S.~Choi, N.~Padmanaban, and G.~Wetzstein, \enquote{Neural holography
  with camera-in-the-loop training,} {\protect\JournalTitle{ACM Transactions on
  Graphics (TOG)}} \textbf{39}, 1--14 (2020).

\bibitem{chakravarthula2020learned}
P.~Chakravarthula, E.~Tseng, T.~Srivastava, H.~Fuchs, and F.~Heide,
  \enquote{Learned hardware-in-the-loop phase retrieval for holographic
  near-eye displays,} {\protect\JournalTitle{ACM Transactions on Graphics
  (TOG)}} \textbf{39}, 1--18 (2020).

\bibitem{li2019progress}
R.~Li and L.~Cao, \enquote{Progress in phase calibration for liquid crystal
  spatial light modulators,} {\protect\JournalTitle{Applied Sciences}}
  \textbf{9}, 2012 (2019).

\bibitem{wu2021high}
J.~Wu, K.~Liu, X.~Sui, and L.~Cao, \enquote{High-speed computer-generated
  holography using an autoencoder-based deep neural network,}
  {\protect\JournalTitle{Optics Letters}} \textbf{46}, 2908--2911 (2021).

\bibitem{zhao2017multi}
T.~Zhao, J.~Liu, X.~Duan, Q.~Gao, J.~Duan, X.~Li, Y.~Wang, W.~Wu, and R.~Zhang,
  \enquote{Multi-region phase calibration of liquid crystal slm for holographic
  display,} {\protect\JournalTitle{Applied optics}} \textbf{56}, 6168--6174
  (2017).

\bibitem{zhang2017calibration}
B.~Zhang, Y.~Chen, and R.~Feng, \enquote{A calibration method for phase-only
  spatial light modulator,} in \emph{2017 Progress In Electromagnetics Research
  Symposium-Spring (PIERS),}  (IEEE, 2017), pp. 133--135.

\bibitem{krasin2021holographic}
G.~Krasin, N.~Stsepuro, I.~Gritsenko, and M.~Kovalev, \enquote{Holographic
  method for precise measurement of wavefront aberrations,} in
  \emph{Holography: Advances and Modern Trends VII,}  vol. 11774 (International
  Society for Optics and Photonics, 2021), p. 1177407.

\bibitem{xun2004phase}
X.~Xun and R.~W. Cohn, \enquote{Phase calibration of spatially nonuniform
  spatial light modulators,} {\protect\JournalTitle{Applied optics}}
  \textbf{43}, 6400--6406 (2004).

\bibitem{vorba2014line}
J.~Vorba, O.~Karl{\'\i}k, M.~{\v{S}}ik, T.~Ritschel, and J.~K{\v{r}}iv{\'a}nek,
  \enquote{On-line learning of parametric mixture models for light transport
  simulation,} {\protect\JournalTitle{ACM Transactions on Graphics (TOG)}}
  \textbf{33}, 1--11 (2014).

\bibitem{ayoub2020review}
M.~Ayoub, \enquote{A review on light transport algorithms and simulation tools
  to model daylighting inside buildings,} {\protect\JournalTitle{Solar Energy}}
  \textbf{198}, 623--642 (2020).

\bibitem{jonsson2020multi}
J.~J{\"o}nsson and E.~Berrocal, \enquote{Multi-scattering software: part i:
  online accelerated monte carlo simulation of light transport through
  scattering media,} {\protect\JournalTitle{Optics Express}} \textbf{28},
  37612--37638 (2020).

\bibitem{reza2019phasor}
S.~A. Reza, M.~La~Manna, S.~Bauer, and A.~Velten, \enquote{Phasor field waves:
  A huygens-like light transport model for non-line-of-sight imaging
  applications,} {\protect\JournalTitle{Optics express}} \textbf{27},
  29380--29400 (2019).

\bibitem{rittig2021neural}
T.~Rittig, D.~Sumin, V.~Babaei, P.~Didyk, A.~Voloboy, A.~Wilkie, B.~Bickel,
  K.~Myszkowski, T.~Weyrich, and J.~K{\v{r}}iv{\'a}nek, \enquote{Neural
  acceleration of scattering-aware color 3d printing,} in \emph{Computer
  Graphics Forum,}  vol.~40 (Wiley Online Library, 2021), pp. 205--219.

\bibitem{corbellini2014connecting}
G.~Corbellini, K.~Aksit, S.~Schmid, S.~Mangold, and T.~R. Gross,
  \enquote{Connecting networks of toys and smartphones with visible light
  communication,} {\protect\JournalTitle{IEEE Communications Magazine}}
  \textbf{52}, 72--78 (2014).

\bibitem{jang2020design}
C.~Jang, O.~Mercier, K.~Bang, G.~Li, Y.~Zhao, and D.~Lanman, \enquote{Design
  and fabrication of freeform holographic optical elements,}
  {\protect\JournalTitle{ACM Transactions on Graphics (TOG)}} \textbf{39},
  1--15 (2020).

\bibitem{akcsit2020patch}
K.~Ak{\c{s}}it, \enquote{Patch scanning displays: spatiotemporal enhancement
  for displays,} {\protect\JournalTitle{Optics express}} \textbf{28},
  2107--2121 (2020).

\bibitem{chakravarthula2018focusar}
P.~Chakravarthula, D.~Dunn, K.~Ak{\c{s}}it, and H.~Fuchs, \enquote{Focusar:
  Auto-focus augmented reality eyeglasses for both real world and virtual
  imagery,} {\protect\JournalTitle{IEEE transactions on visualization and
  computer graphics}} \textbf{24}, 2906--2916 (2018).

\bibitem{li2020optical}
R.~Li, E.~Whitmire, M.~Stengel, B.~Boudaoud, J.~Kautz, D.~Luebke, S.~Patel, and
  K.~Ak{\c{s}}it, \enquote{Optical gaze tracking with spatially-sparse
  single-pixel detectors,} in \emph{2020 IEEE International Symposium on Mixed
  and Augmented Reality (ISMAR),}  (IEEE, 2020), pp. 117--126.

\bibitem{aydindougan2021applications}
G.~Ayd{\i}ndo{\u{g}}an, K.~Kavakl{\i}, A.~{\c{S}}ahin, P.~Artal, and
  H.~{\"U}rey, \enquote{Applications of augmented reality in ophthalmology,}
  {\protect\JournalTitle{Biomedical optics express}} \textbf{12}, 511--538
  (2021).

\bibitem{heurtley1973scalar}
J.~C. Heurtley, \enquote{Scalar rayleigh--sommerfeld and kirchhoff diffraction
  integrals: a comparison of exact evaluations for axial points,}
  {\protect\JournalTitle{JOSA}} \textbf{63}, 1003--1008 (1973).

\bibitem{goodman2005introduction}
J.~W. Goodman, \enquote{Introduction to fourier optics, roberts \& co,}
  {\protect\JournalTitle{Publishers, Englewood, Colorado}}  (2005).

\bibitem{matsushima2009band}
K.~Matsushima and T.~Shimobaba, \enquote{Band-limited angular spectrum method
  for numerical simulation of free-space propagation in far and near fields,}
  {\protect\JournalTitle{Optics express}} \textbf{17}, 19662--19673 (2009).

\bibitem{zhang2020band}
W.~Zhang, H.~Zhang, and G.~Jin, \enquote{Band-extended angular spectrum method
  for accurate diffraction calculation in a wide propagation range,}
  {\protect\JournalTitle{Optics Letters}} \textbf{45}, 1543--1546 (2020).

\bibitem{zhang2020adaptive}
W.~Zhang, H.~Zhang, and G.~Jin, \enquote{Adaptive-sampling angular spectrum
  method with full utilization of space-bandwidth product,}
  {\protect\JournalTitle{Optics Letters}} \textbf{45}, 4416--4419 (2020).

\bibitem{sypek1995light}
M.~Sypek, \enquote{Light propagation in the fresnel region. new numerical
  approach,} {\protect\JournalTitle{Optics communications}} \textbf{116},
  43--48 (1995).

\bibitem{kaan_aksit_2021_5136406}
K.~Akşit, A.~S. Karadeniz, P.~Chakravarthula, W.~Yujie, K.~Kavaklı, Y.~Itoh,
  and D.~R. Walton, \enquote{Odak 0.1.9),}
  https://doi.org/10.5281/zenodo.5526684 (2021).

\bibitem{paszke2017automatic}
A.~Paszke, S.~Gross, S.~Chintala, G.~Chanan, E.~Yang, Z.~DeVito, Z.~Lin,
  A.~Desmaison, L.~Antiga, and A.~Lerer, \enquote{Automatic differentiation in
  pytorch,} in \emph{NIPS 2017 Workshop on Autodiff,}  (2017).

\bibitem{kingma2014adam}
D.~P. Kingma and J.~Ba, \enquote{Adam: A method for stochastic optimization,}
  {\protect\JournalTitle{arXiv preprint arXiv:1412.6980}}  (2014).

\bibitem{loshchilov2017decoupled}
I.~Loshchilov and F.~Hutter, \enquote{Decoupled weight decay regularization,}
  {\protect\JournalTitle{arXiv preprint arXiv:1711.05101}}  (2017).

\bibitem{Ignatov_2018_ECCV_Workshops}
A.~Ignatov, R.~Timofte, T.~V. Vu, T.~M. Luu, T.~X. Pham, C.~V. Nguyen, Y.~Kim,
  J.-S. Choi, M.~Kim, J.~Huang, J.~Ran, C.~Xing, X.~Zhou, P.~Zhu, M.~Geng,
  Y.~Li, E.~Agustsson, S.~Gu, L.~V. Gool, E.~de~Stoutz, N.~Kobyshev, K.~Nie,
  Y.~Zhao, G.~Li, T.~Tong, Q.~Gao, L.~Hanwen, P.~N. Michelini, Z.~Dan,
  H.~Fengshuo, Z.~Hui, X.~Wang, L.~Deng, R.~Meng, J.~Qin, Y.~Shi, W.~Wen,
  L.~Lin, R.~Feng, S.~Wu, C.~Dong, Y.~Qiao, S.~Vasu, N.~T. Madam, P.~Kandula,
  A.~N. Rajagopalan, J.~Liu, and C.~Jung, \enquote{Pirm challenge on perceptual
  image enhancement on smartphones: report,} in \emph{European Conference on
  Computer Vision (ECCV) Workshops,}  (2019).

\bibitem{kavakli_2021}
K.~Kavakl{\i}, H.~Urey, and K.~Ak\c{s}it, \enquote{Phase-only holograms and
  captured photographs,}  (2021).

\bibitem{choi2021optimizing}
S.~Choi, J.~Kim, Y.~Peng, and G.~Wetzstein, \enquote{Optimizing image quality
  for holographic near-eye displays with michelson holography,}
  {\protect\JournalTitle{Optica}} \textbf{8}, 143--146 (2021).

\bibitem{maimone2017holographic}
A.~Maimone, A.~Georgiou, and J.~S. Kollin, \enquote{Holographic near-eye
  displays for virtual and augmented reality,} {\protect\JournalTitle{ACM
  Transactions on Graphics (Tog)}} \textbf{36}, 1--16 (2017).

\end{thebibliography}
\end{document}